\documentclass[11pt,twoside]{article}

\usepackage{asp2006}
\usepackage{epsf}
\usepackage{epsfig}
\usepackage{lscape}

\begin{document}
\title{Sociology of Modern Cosmology}   
\author{Mart\'\i n L\'opez-Corredoira$^1$}   
\affil{$^1$ Instituto de Astrof\'\i sica de Canarias, La 
Laguna, Tenerife, Spain}    

\begin{abstract} 
Certain results of observational cosmology cast critical doubt on the
foundations of standard cosmology but leave most cosmologists untroubled.
Alternative cosmological models that differ from the Big Bang have been
published and defended by heterodox scientists; however, most cosmologists 
do not heed these. This may be because standard theory is correct and
all other ideas and criticisms are incorrect, but it is also to a great
extent due to  sociological phenomena such as the ``snowball effect'' or
``groupthink''. We might wonder whether cosmology, the study of the Universe
as a whole, is a science like other branches of physics or just a dominant ideology.
\end{abstract}



\section{Standard and alternative cosmological models}

The standard (``Big Bang'') model of cosmology  gives us a representation of 
a Cosmos whose dynamics is dominated by gravity (from
gen. relativity), with a finite time, homogeneous 
on large scales, expanding, a hot initial state, together with 
other elements necessary to avoid certain inconsistencies with the observations 
(inflation, non-baryonic dark matter, dark energy, etc.). Although the Big Bang is the most commonly 
accepted theory, it is not the only possible representation of the Cosmos. 
In the last 80 years---such is the short history of this branch of science called 
cosmology---there have been plenty of other proposals.

Among the alternative proposals, there are models with variations on the standard model
although with the same general idea: models with different gravity laws and no need of 
dark matter [e.g., ``Modified Newtonian Dynamics'' (MOND), Sanders \& McGaugh 2002]; 
fractal Universes (Gabrielli et al.\ 2005); cold initial state instead of a hot Big Bang 
(Layzer 1990); variable physical constants; textures instead of inflation; etc.
There are other models that propose a very different scenario with respect to the
standard model: the Quasi-Steady-State Model (Hoyle et al.\ 2000) assumes an eternal time, 
expanding Universe with a superposed cycle of smaller amplitude expansions/contractions, 
large-scale homogeneity, the continuous creation of matter; 
Plasma Cosmology (Lerner 1991) assumes
eternal time, predominance of electromagnetic forces instead of gravity. 
And there are models describing a Universe which is
static, euclidean, infinite space; eternal time models, giving some exotic 
proposals to explain the redshifts of the galaxies within complex gravity theories:
chronometric cosmology (Segal 1976), curvature cosmology (Crawford 2006), 
wave system cosmology (Andrews 1999), negative pressure cosmology (Hawkins 1993), etc.

All models have gaps and caveats to explain some data derived from the observations.
The Big Bang has a lot of problems and aspects which do not work properly or are not 
totally understood yet (see review at L\'opez-Corredoira 2003). The expansion itself has no 
direct proof (nobody has directly observed a galaxy to recede); the most direct argument in
favour of the expansion is the redshift of the galaxies, but the redshift also has other explanations 
than expansion. Other tests on the expansion are dependent on the evolution of the galaxies
or other assumptions. Microwave Background Radiation, light element
abundances, large scale structure formation also have alternative explanations.
And there are problems remaining to be solved: correlations of galaxies and QSOs with different 
redshift; concerning microwave anisotropies (L\'opez-Corredoira 2007, subsect. 5) there are questions such as the 
octopole/quadrupole alignment, non-gaussianity, the 
insufficient lens effect of clusters, etc.; no metallicity evolution (too quick!) observed; failed 
predictions for some elements (lithium and others); observed large-scale structures larger 
than predicted; many problems associated with dark matter (especially on galactic scales), 
etc., etc.
 
Of course, if the Big Bang model has problems, the alternative proposals have their own share of 
 them too, and their problems are sometimes more severe, perhaps because these theories
are not as developed and polished as the standard model. For the expansion, either
they take it as fact, so they need speculative elements to argue that there was no 
beginning of the Universe (e.g., continuous creation of matter in the Quasi-Steady-State
model) or an alternative explanation for the redshift of the galaxies. 
Microwave Background Radiation has alternative explanations different from the Big Bang, 
but with some ad hoc elements (e.g., whiskers to thermalize 
stellar radiation in the Quasi-Steady-State Model) without direct proofs. Also, light element abundances require very 
old populations that have not yet been observed. 

Indeed, alternative models like the Quasi-Steady-State one are not doing 
anything different from the standard model. Its modern version (Hoyle et al. 2000) 
is able to explain most of the difficulties that the previous version of the model (Steady-State) had: 
the existence of young galaxies at high redshift, the distribution of radio 
sources, microwave background radiation, etc. They introduce ad hoc elements without
observational support in the same way that the Big Bang introduces ad hoc 
non-baryonic dark matter, dark energy, inflaton, etc. 
Why, then, are the different theories accepted/rejected with different criteria?

The number of independent measurements relevant to
current cosmology and the number of free parameters 
of the theory are of the same order (Disney 2007):
the ``Big Bang'' was in the '50s a theory with three or four free parameters to 
fit the few numbers of observational cosmological (basically, Hubble's constant 
and the helium abundance), and the increase of cosmological information from
observations, with the CMBR anisotropies\footnote{Regarding 
CMBR anisotropies, the power spectrum is just a curve with two or three clear peaks 
that could be parameterized with $\sim 10$ parameters 
(three parameters/peak: central position, width, height). If we allow certain
range or errors [each peak has important relative error bars, which are very large in the
2nd, 3rd and beyond (indeed, after the 3rd peak the noise dominates)], it is possible
to parameterize a curve like this with somewhat fewer parameters within the errors.
Standard concordance cosmology reproduces the curve with six parameters (there are
indeed $\sim 20$ parameters; but the most important ones are six in number; the rest of them produce small
dependence), with some problems to reproduce the very large scale fluctuations. 
Nonetheless, there also other papers which reproduce the same WMAP data with 
totally different cosmologies with a similar number of free parameters: e.g., Narlikar et al. (2003), 
McGaugh (2004). The fact that different cosmologies with different elements can fit the same data 
(with a similar number of free parameters to fit) indicates that the number of independent numbers 
in the information provided by WMAP data is comparable to the number of free parameters in any of the 
theories.} and others, has been accompanied by an increase in 
free parameters and patches (dark matter, dark energy, inflation) 
in the models to fit those new numbers, until becoming today a theory with
around 20 free parameters (apart from the initial conditions and
other boundary conditions introduced in the simulations to reproduce
certain structures of the Universe). The independent cosmological
numbers extracted from observations are of the same order.
Even so, there are some numbers which cannot be fitted.

The development of modern Cosmology is somewhat similar to the 
development of the epicyclic Ptolemaic theory. However, in this race to build more 
and more epicycles, Big Bang model is allowed to make ad hoc corrections and add more and
more free parameters to the theory to solve the problems
which it finds in its way, but the alternative models are rejected when the gaps
or inconsistencies arise and most cosmologists do not heed their ad hoc corrections.
Why the different theories are accepted/rejected with different criteria?

\section{The difficulties in creating alternative models. A sociological/e\-pis\-te\-mo\-lo\-gi\-cal
model on how Modern Cosmology works}

In my opinion, alternative models are not rejected because they are not potentially
competitive but because they 
they have great difficulties in advancing in their research against the mainstream. 
A small number of scientists cannot compete with the huge mass of cosmologists 
dedicated to polishing and refining the standard theory.
The present-day methodology of research in Cosmology does not favour the exploration
of new ideas. The standard theory in Cosmology became dominant because 
it could explain more phenomena than the alternative ideas, but it is possible that 
partial successes have propitiated the compromise with a general 
view which is misguided and does not let other ideas advance that might be closer 
to a correcter description of the Universe. 

\subsection{Methodology of science}

Basically, there are two different methodologies to study Nature, both inherited from
different ways of thinking in ancient Greece:

\begin{description}

\item[The mathematical deductive method:]
This is the method thought by Py\-tha\-go\-ras and Plato. The pure relations of numbers in Arithmetic and Geometry
are the immutable reality behind changing appearances in the world of the senses.
We cannot reach the truth through observation with the senses, but only through pure
reason, which may investigate the abstract mathematical forms that govern the world.
An example within modern science might be Einstein's general relativity, which was
posited from aesthetic and/or rational principles in a time in which observational
data did not require a new gravity theory. Physics and cosmology nowadays are
partially Pythagorian when a theory is created before the observations.

\item[The empirical inductive method:] This is the method thought 
by Anaxagoras of how to know Nature. Aristotle uses both inductive and deductive methods, and he 
says that ``the mathematical method is not the method of the physicists, because
Nature, perhaps all, involves matter'' (Metaphysics, book II). Matter and not numbers or mathematics.
Nature should be known through observations and extrapolations of them.
The empiricism of Galileo Galilei might be an example within modern science, although all 
scientists, even Galileo, are also partly Pythagorean.  
In my opinion, Cosmology should be derived empirically by first taking the data and then interpreting
them from all possible theoretical solutions.
As Sherlock Holmes said:
``It is a capital mistake to theorize before you have all the 
evidence'' (\textit{A Study in Scarlet}); ``Before one has data, one begins to twist facts to 
suit theories instead of theories to suit data'' (\textit{A Scandal in Bohemia}) [cited by
Burbidge 2006].

\end{description}

Some astrophysicists closer to the observations than the theory usually complain
about the lack of an empirical approach in Cosmology. For instance, G\'erard de 
Vaucouleurs (1918--95), known for his extragalactic surveys and Hubble constant
measurements, said that there are  ``parallelisms between modern 
cosmology and medieval scholasticism. (...) Above all 
I am concerned by an apparent loss of contact with empirical evidence and observational facts, 
and, worse, by a deliberate refusal on the part of some theorists to accept such results 
when they appear to be in conflict with some of the present oversimplified and therefore 
intellectually appealing theories of the universe'' (de Vaucouleurs 1970).
There is, however, an epistemological optimism encouraging the belief that the successful theories
are successful because they reflect  reality in Nature. This might be true in certain branches
of science but not in those areas close to metaphysical speculations such as cosmology, where
the scientific method is something like:

\begin{quotation}
''--- Given a theory A self-called orthodox or standard, and a non-orthodox or 
non-standard theory B. If the observations achieve what was predicted by the 
theory A and not by the theory B, this implies a large success to the theory A, 
something which must be divulged immediately to the all-important mass 
media. This means that there are no doubts that theory A is the right one. 
Theory B is wrong; one must forget this theory and, therefore, any further 
research directed to it must be blocked (putting obstacles in the way of 
publication, and giving no time for telescopes, etc.).

--- If the observations achieve what was predicted by theory B rather than by 
theory A, this means nothing. Science is very complex and before taking a 
position we must think further about the matter and make further tests. It is 
probable that the observer of such had a failure at some point; further 
observations are needed (and it will be difficult to make further observations 
because we are not going to allow the use of telescopes to re-test such a 
stupid theory as theory B). Who knows! Perhaps the observed thing is due to 
effect `So-and-so', of course; perhaps they have not corrected the data from 
this effect, about which we know nothing. Everything is so complex. We must 
be sure before we can say something about which theory is correct. 
Furthermore, by adding some new aspects in the theory A surely it can 
also predict the observations, and, since we have an army of theoreticians 
ready to put in patches and discover new effects, in less 
than three months we will have a new theory A (albeit with some changes) 
which will agree the data. In any case, while in troubled waters, and as long 
as we do not clarify the question, theory A remains. Perhaps, as was said by 
Halton Arp, the informal saying `to make extraordinary changes one requires 
extraordinary evidence' really means `to make personally disadvantageous 
changes no evidence is extraordinary enough'.'' (L\'opez-Corredoira 2008)
\end{quotation}

Halton C. Arp (1927-- ), 
reputedly heterodox observational cosmologist known through
his proposal of non-cosmological redshifts (L\'opez-Corredoira 2003, \S 2.8) 
would point out: ``Of course, if one ignores contradictory observations, 
one can claim to have an 'elegant' or 'robust' theory. But it isn't 
science.'' (Arp \& Block 1991)

\subsection{The snowball effect}

The alternative models try to compete with the standard model, but  cumulative
inertia gives a clear advantage to the standard model:
``The snowball effect arising from the social dynamics of research funding drove more 
researchers into the Standard Cosmology fold and contributed to the drying out 
of alternative ideas.'' (Narlikar \& Padmanabhan 2001). 
It is not strange that Jayant V. Narlikar (1938-- ), 
one of the creators of the Quasi-Steady-State model,
links the lack of social success of his theory to how social dynamics works. Anyway, what
he claims is basically correct and applicable to most speculative sciences.
Another creator of the Quasi-Steady-State, Geoffrey R. Burbidge (1925-- ), 
does not have a higher opinion:

\begin{quotation}
``Let me start on a somewhat pessimistic note. We all know that new ideas and revolutions 
in science in general come from the younger generation, who look critically at the contemporary 
schemes, and having absorbed the new evidence, overthrow the old views. This, in general, is the 
way that science advances. However, in modern astronomy and cosmology, at present, this is 
emphatically not the case. Over the last decade or more, the vast majority of the younger 
astronomers have been conformists in the extreme, passionately believing what their leaders 
have told them, particularly in cosmology. In the modern era the reasons for this are even 
stronger than they were in the past. To obtain an academic position, to obtain tenure, to be 
successful in obtaining research funds, and to obtain observing time on major telescopes, it 
is necessary to conform.'' (G. R. Burbidge 1997)
\end{quotation}

The snowball effect, also called Matthew effect (Merton 1968),
is to a certain extent present in the social dynamics of  Cosmology.
It is a feedback ball: the more successful the standard theory is, the more money and
scientists are dedicated to work on it, and therefore the higher the number of observations
that can be explained ad hoc, and that lead the theory to be considered more successful.
However, not everything is a social
construct (as some postmodernists claim); the microwave background radiation, the redshift of
galaxies, etc., are real facts and they also have  weight in the credibility of the
standard model.

\subsection{Psychological profile of cosmologists}

There are two main psychological profiles of cosmologists, with gradations of grey between them:

\begin{description}

\item[Heterodox:] possessed by the complex of unappreciated genius, 
too much ``ego'', normally working alone/individually or in very small groups,
creative, intelligent, non-conformist. His/her (mostly males)
dream is to create a new paradigm in science which completely changes  our view 
of the Universe. Many of them try to demonstrate that Einstein was wrong, 
maybe because he is the symbol of genius and defeating his theory would mean that
they are geniuses above Einstein. Most of them are crackpots.

\item[Orthodox:] dominated by the groupthink, following a leader's opinion
as in the ``Naked king'' tale\footnote{After I offered my talk on this paper
in the present conference ``Cosmology across Cultures'', Joel Primack stood and
shouted annoyedly ``You are ignorant!, you are ignorant!...''
several times. This is a good example of ``Naked king''-type behaviour,
also very common among the defenders of contemporary art and priests of many 
religions: if you do not accept my view, it is because you are an ignorant. 
I do not know whether the present paper contains elements that deserve to be called 
``ignorant''. Primack  offered us an excellent presentation with magnificent 
simulations and special effects on how our Universe formed and evolved, emphasizing that
the Cold Dark Matter + Dark Energy model  accurately ``predicts''  the observations;
and his wife, the artist, writer and philosopher Nancy E. Abrams, offered in her talk 
a guideline to our lives claiming that her husband's hypothesis
is something eternal since, as in the case of  Newtonian gravitation with
respect to general relativity, scientific theories get broader but are not
demonstrated to be false. I do not agree Primack's concept of 
``prediction''  when we are doing ad hoc fits, and I do not agree
that theories are eternal (Ptolemaic geocentric astronomy, phlogiston theory,
caloric theory, Newtonian optics, ether, etc. were proven to be wrong ideas). As in most Hollywood historical 
films, I see very good special effects but not very accurate descriptions of
history. Nonetheless, I do not think that the show by Primack+Abrams came from 
ignorant people but from different standpoints.}, good workers performing monotonous 
tasks without ideas in large groups, specialists in a small field which they know 
very well, conformist, domestic. His/her dream is getting a permanent position 
at an university or research center, to be leader of a project, to do astropolitics
(see L\'opez-Corredoira 2008). Most of them are like sheep (or geese\footnote{In 
Hoyle et al. (2000), a serious and technical book about Cosmology,
a picture was inserted in which a row of geese are turning around a corner all 
in the same way, with the following ironic comment:
``This is our view of the conformist approach to the standard
(hot big bang) cosmology. We have resisted the temptation to name some of the leading
geese''.}), some of them with vocation of shepherds too.

\end{description}

The sociological reasons for favouring  orthodox proposals might be related to the
preference of domesticity in our civilization. Sheep rather than crackpots are preferred.
Finding a promising change of paradigm closer to the truth among thousands of crazy 
proposals is very difficult; in orthodoxy, although  absolute truth is not guaranteed, 
at least a consensus version of the truth is offered. 

\section{Limits of Cosmology}

\begin{quotation}
And we would pretend to understand everything about cosmology, which concerns the whole 
Universe? We are not even ready to start to do that. All that we can do is to enter in the field 
of speculations. So far as I am concerned, I would not comment myself on any cosmological theory, 
on the so-called `standard theory' less on many others. Actually, I would like to leave the door 
wide open.  (Pecker 1997)
\end{quotation}

I agree Jean-Claude Pecker (1923-- ), 
another classical heterodox dissident cosmologist.
Before wondering which is the true model of cosmology, we must wonder whether we are
in a condition to create a theory on the genesis (or non-genesis) and evolution of the
whole Universe, whether the psychological/so\-cio\-lo\-gi\-cal conditions of the cosmologists
are or are not a weightier factor than the observations of Nature.

There are limits to Cosmology because we are finite human beings limited by our experiences
and circumstances, not mini-gods able to read the mind of a God who played maths with
the Universe, as some Pythagoreans may think. There is a lack of humility in Pythagorism,
or in expressions like ``precision Cosmology''. One of the most reputed 
physicists of the former Soviet Union,
Lev Lavidovich Landau (1908--68),
said: ``Cosmologists are often in error, but never in doubt''.
Great old masters, even the creators of the standard model, were cautious
in their assertions. Edwin P. Hubble (1889--1953) throughout his life doubted the reality 
of the expansion of the Universe. Willem de Sitter (1872-1934) claimed: 
``It should not be forgotten that all this talk about the universe 
involves a tremendous extrapolation, which is a very dangerous operation'' (de Sitter 1931).
This scepticism is sane since ``all cautions are too few'' (Spanish proverb).
It is not a question of substituting one model for another, since it would be the ``same
dog with different collar'' (another Spanish proverb) but of realizing  the limits
of Cosmology as a science. Before understanding the Universe, we must understand the galaxies.

Rutherford (1871--1937) said ``Don't let me hear anyone use the word `Universe' 
in my department.''. With the same style, the astrophysicist 
Mike Disney (1937-- ) 
dared to claim: ``The word `cosmologist' should be expunged from 
the scientific dictionary and returned to the priesthood where it properly belongs.'' 
(Disney 2000). Words of an old-style scepticism. Nowadays, the young bloods of precision
Cosmology do not care about those statements and  proudly claim  that people in the past did not know what they know.
Cosmologists with no indication of doubt and an amazing sense of security who
dissert on topics of high speculation. Of course, science advances, and 
Cosmology advances in the amount of data and epicycle-like 
patches in the theory to fit the data, but the great questions remain 
almost unchanged. Many wise men have already deliberated on cosmology over 
long time, without reaching a definitive solution.
Do we live in a fortunate golden age of Cosmology that allows us, 
thanks to our technical advances and our trained researchers, 
to answer questions on eternity, finiteness of the Universe, etc.? 
We could reply as the  XIXth century German philosopher Schopenhauer
did with the Know-alls of his time:

\begin{quotation}
``Every 30 years, a new generation of talkative candid persons, ignorant
of everything, want to devour summarily and hastily the results of human
knowledge accumulated over centuries, and immediately they think themselves
more skillful than the whole past.''
\end{quotation}

\acknowledgements 
Thanks are given to Terry J. Mahoney (IAC, Tenerife, Spain) for
proof-reading this paper.


\end{document}